\documentclass[aip,jmp,amsmath,amssymb,reprint]{revtex4-1}

\usepackage{empheq}
\usepackage{graphicx}
\usepackage{dcolumn}
\usepackage{bm}

\usepackage{cprotect}
\usepackage{soul,color}
\usepackage{etoolbox}
\usepackage{floatrow}

\begin{document}
\title{Warped phase coherence: an empirical synchronization measure combining phase and amplitude information}
\author{Ludovico Minati}
\altaffiliation{Author to whom correspondence should be addressed. Electronic addresses: minati.l.aa@m.titech.ac.jp and lminati@ieee.org. Tel.: +39 335 486 670. URL: http://www.lminati.it.}
\affiliation{Tokyo Tech World Research Hub Initiative - Institute of Innovative Research, Tokyo Institute of Technology, Yokohama 226-8503, Japan}
\affiliation{Complex Systems Theory Department, Institute of Nuclear Physics - Polish Academy of Sciences (IFJ-PAN), 31-342 Krak\'ow, Poland}
\affiliation{Center for Mind/Brain Sciences (CIMeC), University of Trento, 38123 Trento, Italy}
\author{Natsue Yoshimura}
\affiliation{FIRST - Institute of Innovative Research, Tokyo Institute of Technology, Yokohama 226-8503, Japan}
\affiliation{PRESTO, JST, 332-0012 Saitama, Japan}
\author{Mattia Frasca}
\affiliation{Department of Electrical Electronic and Computer Engineering (DIEEI), University of Catania, 95131 Catania, Italy}
\author{Stanis\l aw Dro\.{z}d\.{z}}
\affiliation{Complex Systems Theory Department, Institute of Nuclear Physics - Polish Academy of Sciences (IFJ-PAN), 31-342 Krak\'ow, Poland}
\affiliation{Faculty of Physics, Mathematics and Computer Science, Cracow University of Technology, 31-155 Krak\'ow, Poland}
\author{Yasuharu Koike}
\affiliation{FIRST - Institute of Innovative Research, Tokyo Institute of Technology, Yokohama 226-8503, Japan}
\date{\today}
\begin{abstract}
The entrainment between weakly-coupled nonlinear oscillators, as well as between complex signals such as those representing physiological activity, is frequently assessed in terms of whether a stable relationship is detectable between the instantaneous phases extracted from the measured or simulated time-series via the analytic signal. Here, we demonstrate that adding a possibly complex constant value to this normally null-mean signal has a non-trivial warping effect. Among other consequences, this introduces a level of sensitivity to the amplitude fluctuations and average relative phase. By means of simulations of R\"ossler systems and experiments on single-transistor oscillator networks, it is shown that the resulting coherence measure may have an empirical value in improving the inference of the structural couplings from the dynamics. When tentatively applied to the electroencephalogram recorded while performing imaginary and real movements, this straightforward modification of the phase locking value substantially improved the classification accuracy. Hence, its possible practical relevance in brain-computer and brain-machine interfaces deserves consideration.
\end{abstract}
\maketitle
\begin{quotation}
In between the extremes of complete asynchrony and perfect synchronization between non-linear dynamical systems, complicated trajectories can show associations in some aspects but not others. A frequent observation is the presence of a relatively stable phase relationship between the positions of two systems along their respective orbits, while the sizes, i.e. amplitudes, of these orbits fluctuate more or less independently. We introduce a straightforward algebraic change to the well-known phase locking value and demonstrate how the resulting warping, among other consequences, confers a level of sensitivity to the amplitude fluctuations. We find that the measure thus obtained is potentially useful, for example as regards enhancing the ability of classifying imaginary actions based on short electroencephalogram segments.
\end{quotation}
\section{INTRODUCTION}\label{intro}
Networked nonlinear, possibly chaotic oscillators which are nonidentical by virtue of parametric heterogeneities or structural nonequivalence pervade the natural and physical world. In such systems, the absence of an invariant manifold implies a nontrivial effect of the coupling strength. Entrainment initially ensues in the form of an increasingly stable phase relationship coexisting with incoherent amplitudes, because the energy transfer rate is insufficient to fully overcome the repulsive forces and lock a common trajectory. As coupling is further strengthened, amplitude fluctuations become increasingly entrained and finally a common trajectory is attained, hallmarking complete synchronization.\cite{rosenblum1996,boccaletti2002,boccaletti2018} Though this behavior is generally well-established, even in simple situations such as coupled R\"ossler systems non-trivial routes to synchronization can be observed depending on the topology of the underlying attractors.\cite{osipov2003} The measurement of synchronization remains an area of active investigation, and a variety of approaches are available. On the one hand, measures of phase locking are particularly sensitive to weak couplings, and are accordingly of wide interest. For example, they represent an empirical means of establishing functional associations between brain areas, based on noisy, broadband signals such as the electroencephalogram. However, they completely disregard amplitude information: particularly in the presence of noise, this can be an undesirable limitation, since equal weighting is given to the synchrony of small and large fluctuations.\cite{kovach2017,lepage2017} On the other hand, measures such as cross-correlation, spectral coherence, mutual information, and generalized synchronization include both phase and amplitude information. However, they may be poorly sensitive in situations of partial synchronization where amplitude fluctuations remain largely uncorrelated and may require longer time-series for reliable estimation.\cite{lowet2016,prieto2017} Accordingly, the phase locking value and its variants are often the methods of choice for the design of connectivity-based brain-computer and brain-machine interfaces, also in virtue of their rapid convergence and robustness to non-stationarity.\cite{brunner2016,sburlea2017} A measure primarily indexing phase synchronization but capable of including a variable amount of amplitude information could have considerable practical utility.
\section{FORMULATION}\label{formulation}
\begin{figure}
\begin{center}
\includegraphics[width=\textwidth]{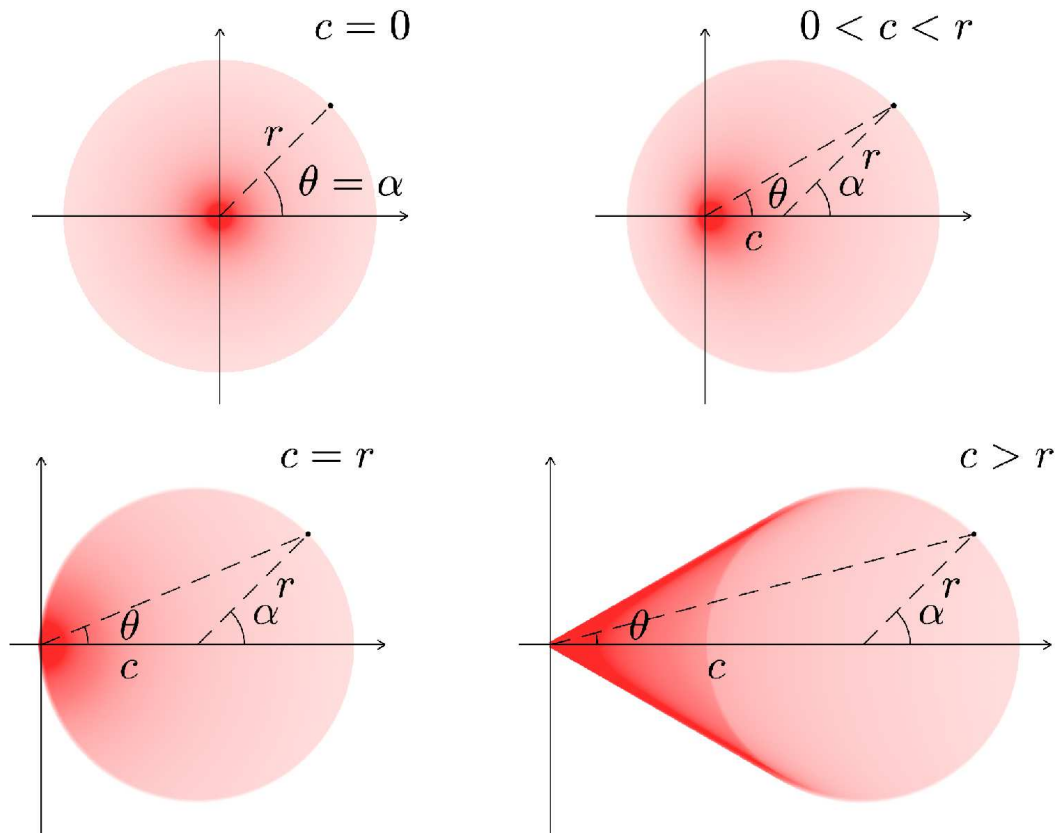}
\end{center}
\caption{Geometry on the complex plane of warping the instantaneous phase $\alpha$ into angle $\theta$ depending on the amplitude $r$ and constant parameter $c$ (here assumed real non-negative) with Eq. (5). Red hue denotes radii between the origin and the circle having radius $r$ centered around $(c,0)$.}
\label{fig:fig1}
\end{figure}
\begin{figure}
\begin{center}
\includegraphics[width=\textwidth]{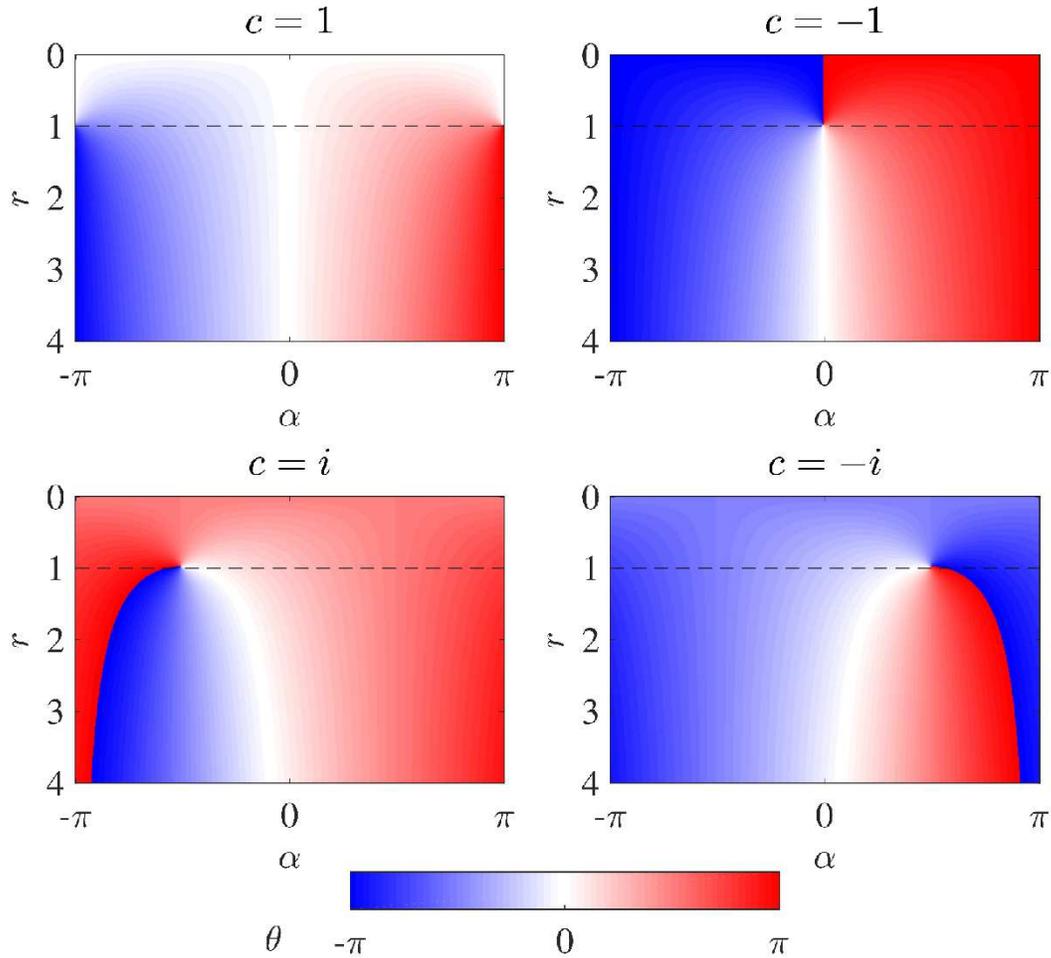}
\end{center}
\caption{Warping the phase $\alpha$ into the angle $\theta$ with Eq. (5), assuming $|c|=1$, visualizing the different effect depending on whether $r<|c|$, $r=|c|$ or $r>|c|$ as also captured in Eqs. (6) and (7).}
\label{fig:fig2}
\end{figure}
\begin{figure}
\begin{center}
\includegraphics[width=\textwidth]{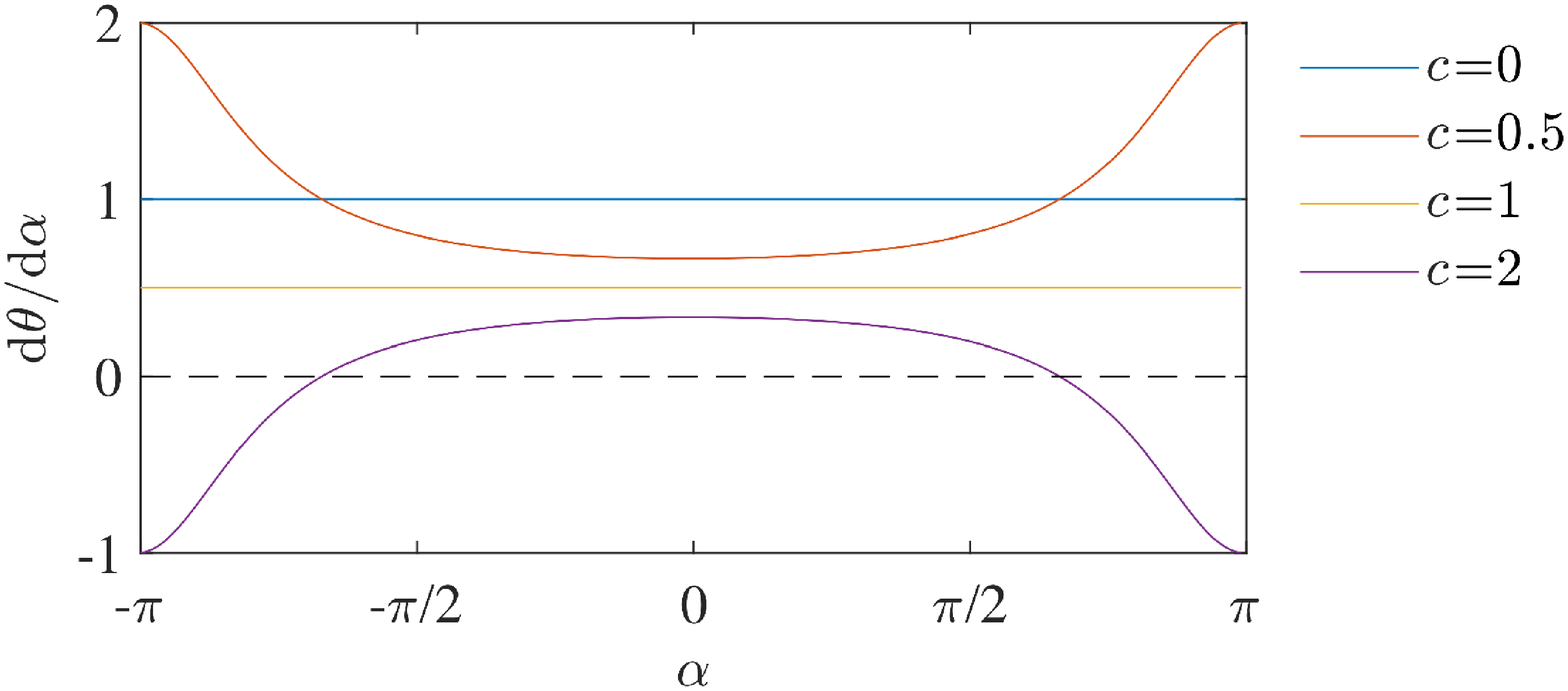}
\end{center}
\caption{Relationship between the phase $\alpha$ and the angle $\theta$ observed as $c$ is increased, corresponding to the four situations in Fig. \ref{fig:fig1} and assuming $r=1$. For $c=2$, the relationship becomes non-monotonic, i.e. $\textrm{d}\theta/\textrm{d}\alpha$ crosses zero. Hence, the angle $\theta$ no longer represents a phase.}
\label{fig:fig3}
\end{figure}
Given a set of $n$ real-valued time-series $s_j(t)$ with $j=1\ldots n$, the first step in quantifying the phase locking is obtaining the corresponding analytic signals. These separate the instantaneous amplitude and phase components $r_j(t)\in[0,\infty)$ and $\alpha_j(t)\in[-\pi,\pi]$,
\begin{equation}
\psi_j(t)=s_j(t)+i\tilde{s}_j(t)=r_j(t)e^{i\alpha_j(t)}\textrm{ ,}
\end{equation}
where $i=\sqrt{-1}$ and $\tilde{s}_j(t)$ is the Hilbert transform of $s_j(t)$
\begin{equation}
\tilde{s}_j(t)=\frac{1}{\pi}\textrm{p.v.}\left[\int_{-\infty}^\infty\frac{s_j(\tau)}{t-\tau}\textrm{d}\tau\right]\textrm{ ,}
\end{equation}
and where $\textrm{p.v.}$ denotes the Cauchy principal value of the integral.\cite{boashash1992} The instantaneous relative phase between a pair of time-series or signals $(j,k)$ can then be obtained with 
\begin{equation}
\Delta\alpha_{jk}(t)=\arg(\psi_j(t)\overline{\psi}_k(t))\textrm{ .}
\end{equation}
From this, the degree of phase synchronization, or phase locking value, is defined as
\begin{equation}
d_{jk}=|\langle e^{i\Delta\alpha_{jk}(t)}\rangle_t|=|\langle e^{i[\alpha_j(t)-\alpha_k(t)]}\rangle_t|\textrm{ .}
\end{equation}
While this parameter fully represents the level of coherence in Kuramoto-like networks, it only provides a partial account of entrainment in systems possessing free amplitude, such as Stuart-Landau and R\"ossler oscillators. This encompasses the near-totality of experimental scenarios. Furthermore, in complex networks of such systems, amplitude dynamics support the onset and maintenance of phase synchronization.\cite{gambuzza2016}\\
Normally, one has $\langle \psi_j(t) \rangle=0$. The present work is a preliminary exploration concerned with the notable effects of adding a possibly complex constant parameter $c$ to the analytic signal, such that
\begin{equation}
\theta_j(t,c)=\arg(\psi_j(t)+c)\textrm{ .}
\end{equation}
By this straightforward operation, the instantaneous phases $\alpha_j(t)$ are warped into angles $\theta_j(t,c)$ which also depend on the amplitudes $r_j(t)$. Unlike in a simple weighting, these are determined by the magnitude $|c|$ as well as by the warping direction $c/|c|=\pm1,\pm i$.\\
First, let us consider that the underlying geometric relationships determine four possible situations on the complex plane, hereby illustrated without loss of generality assuming a uniform distribution $\alpha\in[-\pi,\pi]$ and $c/|c|=1$ (Fig. \ref{fig:fig1}). If $c=0$, there is no warping and $\theta=\alpha$. For $0<c<r$, one still has $\theta\in[-\pi,\pi]$ but the distribution is gradually drawn towards a peaked one, eventually leading again to a uniform distribution in $\theta\in[-\pi/2,\pi/2]$ when $c=r$. For $c>r$, the distribution is morphed towards an increasingly narrow centrally-symmetric one. In the limit $c\rightarrow\infty$, one has $\theta\in[-\tan^{-1}(r/c),\tan^{-1}(r/c)]$.\\
In other words, the instantaneous amplitude $r$ influences the effect of the phase $\alpha$ on the angle $\theta$ in an intricate manner, which can be conceptualized as follows. On the one hand, the surface is ``twisted'', so that if $r\ll|c|$, $\delta\alpha$ is shrunk into $\delta\theta\approx0$, whereas if $r\gg|c|$ one has $\delta\theta\approx\delta\alpha$. As discussed below, when $\theta$ is used as a basis for determining synchronization, this emphasizes the phase coherence of large-amplitude fluctuations. On the other hand, the surface is also ``plucked'', so that for $r\approx|c|$ a large swing is observed in the vicinity of a singularity point at $\alpha=\arg(-c)$, around which the relationship between $\theta$ and $\alpha$ diverges (Fig. \ref{fig:fig2}).\\
Importantly, Eq. (5) implies a possibly non-monotonic relationship between the phase $\alpha$ and the angle $\theta$, namely when $c>r$ (Fig. \ref{fig:fig3}). Hence, after the warping, the angle $\theta$ can no longer be considered as a phase because, assuming a fixed frequency, it may not satisfy the fundamental requirement of increasing (or decreasing) monotonically with time. When $c>r$, the synchronization measure described below and based on this value should then not be considered as properly representing phase synchronization. At most, it only has an empirical value.\cite{pikovsky2001}\\
Another perspective on the warping effect is obtained by considering, again assuming $c/|c|=\pm1$,
\begin{empheq}{align}
\frac{\partial\theta}{\partial\alpha}&=\frac{r^2+cr\cos\alpha}{r^2+2cr\cos\alpha+c^2}\textrm{ and}\\
\frac{\partial\theta}{\partial r}&=\frac{c\sin\alpha}{r^2+2cr\cos\alpha+c^2}\textrm{ .}
\end{empheq}
Here, the equations for $c/|c|=\pm i$ are readily obtained by replacing $\cos\alpha$ with $\sin\alpha$, $\sin\alpha$ with $-\cos\alpha$, and $c$ with $c/i$. While the effect of $r$ on $\partial\theta/\partial\alpha$ is non-trivial and depends on $\alpha$, away from $\alpha=\arg(-c)$ a positive relationship is evident, whereby phase fluctuations are magnified when amplitude is large.\\
Second, a measure of warped phase coherence can be introduced by replacing the phase $\alpha$ with the angle $\theta$ in Eq. (4), as 
\begin{equation}
\hat{w}_{jk}(c)=|\langle e^{i[\theta_j(t,c)-\theta_k(t,c)]}\rangle_t|\textrm{ ,}
\end{equation}
where by definition $d_{jk}=\hat{w}_{jk}(0)$. At a given setting of $c$, $\hat{w}_{jk}(c)$ reflects the level of synchrony as visible after the phase warping. However, as the value of $c$ is changed an undesirable effect is noted: due to the influence of $|c|$ on the distribution of $\theta$, in $|c|\rightarrow\infty$ one has $\hat{w}\rightarrow1$. As depicted in Fig. \ref{fig:fig1}, this is purely because $\theta\rightarrow0$. When appropriate, e.g. for visualization purposes, this issue can be addressed by normalizing, for example according to
\begin{equation}
w_{jk}(c)=\frac{\hat{w}_{jk}(c)-\hat{w}^\prime_{jk}(c)}{1-\hat{w}^\prime_{jk}(c)}\textrm{ ,}
\end{equation}
where 
\begin{equation}
\hat{w}^\prime_{jk}(c)=|\langle e^{i[\theta_j(t,c)-\Omega\left[\theta_k(t,c)]\right]}\rangle_t|\textrm{ .}
\end{equation}
Here, $\Omega\left[\mathbf{x}\right]$ reshuffles $x_u$ according to a random sequence of index $u$. This operation retains the angle distributions $p(\theta_j)$ and $p(\theta_k)$, but alters the distribution corresponding to $p(\theta_j-\theta_k)$. The underlying rationale is that, due to the reshuffling operation which destroys any temporal relationship, $\hat{w}^\prime_{jk}(c)$ approximates the value of $\hat{w}_{jk}(c)$ that would be observed if the signals were completely incoherent, taking into account the angle distributions $p(\theta_j)$ and $p(\theta_k)$. Consequentially, this allows rescaling the coherence values onto a range closer to $[0,1]$. This operation can have complex consequences depending on the distributions. However, for the avoidance of doubt, it is herein introduced only as a potentially convenient form of rescaling. It is not as an essential step in the determination of warped coherence: in fact, this rescaling has the drawback that the shuffling operation in Eq. (10) can introduce random error. In the absence of entrainment, one expects $w_{jk}(c)\approx0$; contrariwise, as exemplified in Section \ref{random}, one can have $w_{jk}(c)>0$ or even $w_{jk}(c)<0$.\\
\section{Random signals}\label{random}
\begin{figure}
\begin{center}
\includegraphics[width=\textwidth]{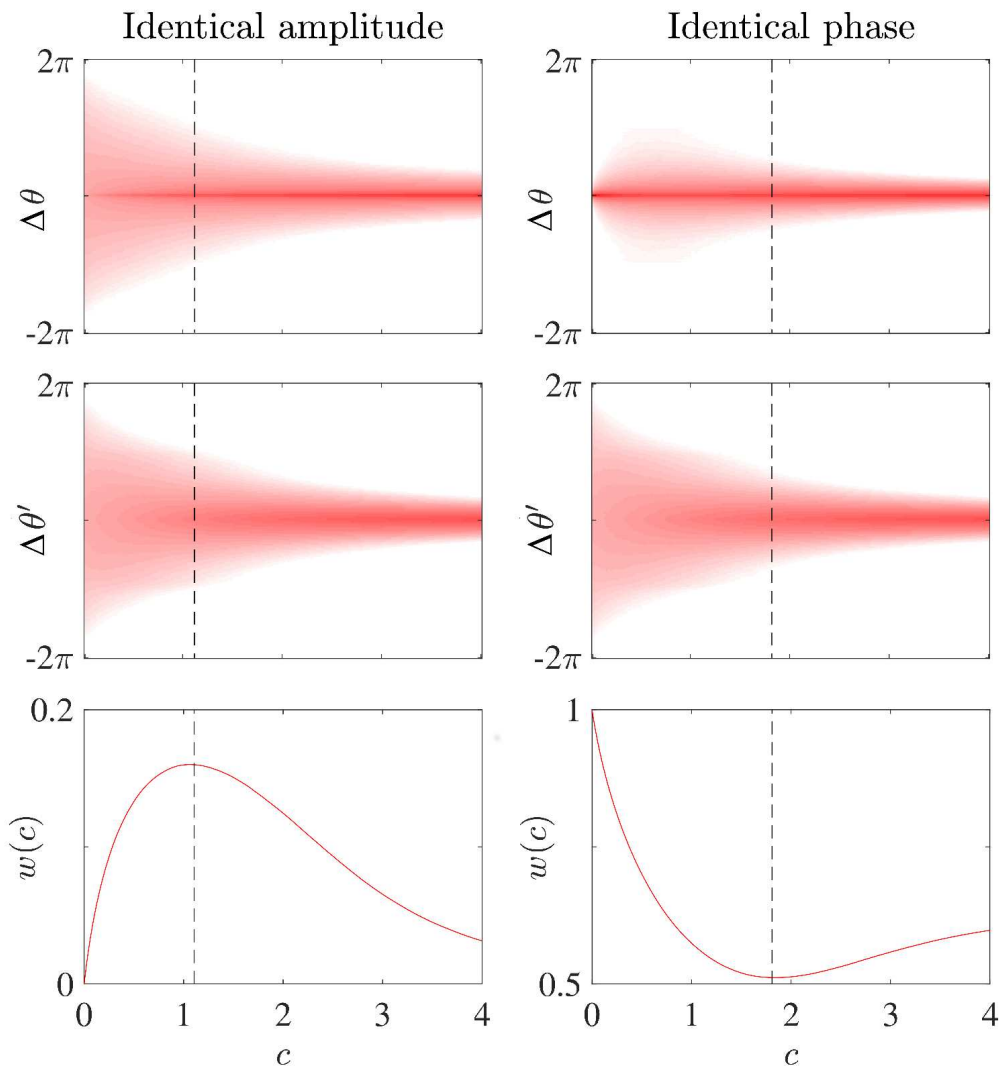}
\end{center}
\caption{Random signals with identical amplitude (left) or phase (right): distributions of relative warped angles before ($\Delta\theta$) and after ($\Delta\theta^\prime$) reshuffling, and resulting normalized warped phase coherence $w(c)$ (dashed lines denote maximum and minimum). Red color intensity proportional to the logarithm of probability density.}
\label{fig:fig4}
\end{figure}
\begin{figure}
\begin{center}
\includegraphics[width=\textwidth]{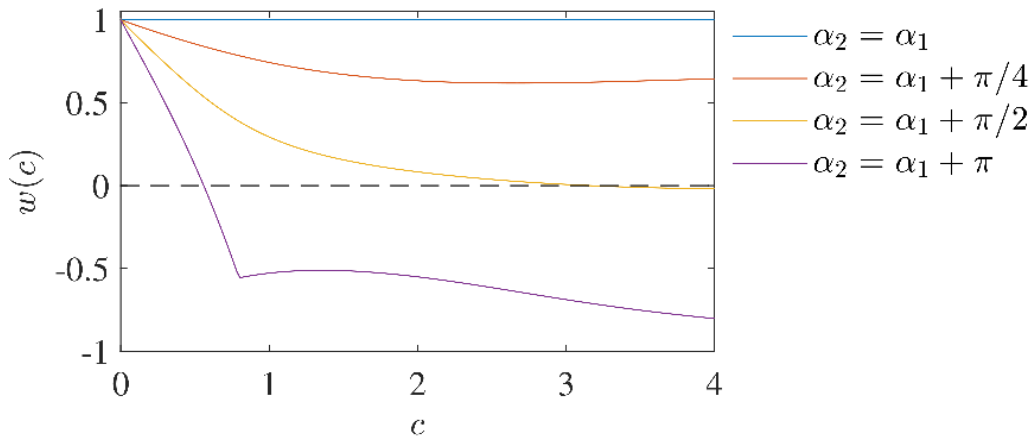}
\end{center}
\caption{Effect of non-zero average relative phase on the normalized warped phase coherence $w(c)$ between two random signals with fully coherent phases and amplitudes.}
\label{fig:fig5}
\end{figure}
The precise effect of warping depends on the phase and amplitude distributions. Nevertheless, the additional sensitivity of warped phase coherence to amplitude synchronization can be initially illustrated via two paradigmatic cases involving pairs of random signals $j=1,2$ having the form $\psi_j=|r_j|e^{i\alpha_j}$, where $r_j$ and $\alpha_j$ are random variables. In one case, the amplitudes are identical but the phases are not, i.e. $r_1=r_2$ and $\alpha_1\neq\alpha_2$, whereas in the other case the converse is true, i.e. $r_1\neq r_2$ and $\alpha_1=\alpha_2$.\\ In both cases, $r_j$ and $\alpha_j$ are respectively drawn from a Gaussian distribution with $\mu=0$ and $\sigma=1$, and from a uniform distribution in $[-\pi,\pi]$. Unless otherwise noted, here and thereafter the signals are rescaled to unitary average amplitude, i.e. $\langle |\psi| \rangle=1$.\\
Let us consider the angle differences for the initial and reshuffled signals as per Eqs. (8) and (10), namely $\Delta\theta=\theta_1(t,c)-\theta_2(t,c)$ and $\Delta\theta^\prime=\theta_1(t,c)-\Omega\left[\theta_2(t,c)\right]$. While $p(\Delta\theta)$ and $p(\Delta\theta^\prime)$ both tend towards Dirac delta functions in $|c|\rightarrow\infty$, they do so along different trajectories: their respective distributions determine the form of $w(c)$. Namely, in the case of identical amplitude but unrelated phases, as expected one has $w(0)=0$, above which the measured coherence increases until $w(1.1)=0.16$ then decreases again. Contrariwise, in the case of identical phase but unrelated amplitudes, as expected one has $w(0)=1$, above which the measured coherence decreases until $w(1.8)=0.51$, then marginally increases again (Fig. \ref{fig:fig4}).\\
For $c>0$, depending on the strength of amplitude entrainment, $w(c)$ can therefore increase or decrease with respect to the amplitude-agnostic observation established by the non-warped value at $w(0)$, effectively corresponding to the phase locking value. It is therefore evident that the effect of warping is fundamentally different from that of recently-proposed amplitude-weighted versions of the phase locking value. In these cases, regardless of the weighting one always has null and unitary expected value when the underlying phases are, respectively, fully incoherent and fully coherent.\cite{kovach2017,lepage2017}\\
Importantly, in the context of two coupled non-linear dynamical systems, the situation of entrained amplitudes but incoherent phases is rather unphysical. That is because, even in presence of generalized synchronization, a functional relation between the instantaneous states must be present to retain coherent amplitudes. However, such scenario can easily arise via modulation mechanisms, for example when two signal sources are such that their instantaneous phases are generated by independent random processes, but their cycle amplitudes are generated by other processes which are entrained. This is empirically observed for neurophysiological signals such as the electroencephalogram, which represent a convoluted ensemble average over a large number of partially-coherent non-linear generators. Indeed, even exotic situations such as preferential phase-amplitude couplings can be observed for such signals, and may index fundamental functions such as memory formation.\cite{canolty2010}\\
On another note, phase synchronization clearly does not necessarily imply the identity of phases, hence the effect of the average relative phase should be considered. The above case with $\alpha_1=\alpha_2$ is not generally representative, but only an initial demonstration of the sensitivity to the amplitudes $r_1, r_2$.\cite{pikovsky2001}  Considering the case $r_1=r_2$ and $\alpha_2=\alpha_1+\beta$ where $\beta\in[-\pi,\pi]$ is a constant parameter, reveals that warped phase coherence is sensitive to the average relative phase, eventually leading to a change in sign. When $\beta=0$, $w(c)=1$ regardless of $c$; however, for increasing $|\beta|$, $w(c)$ gradually decays and, for sufficiently strong warping, one observes $w(c)<0$ when $|\beta|>\pi/2$ (Fig. \ref{fig:fig5}). This is reminiscent of linear correlation and obviously deviates from the canonical definition of phase synchronization, which is by definition agnostic to relative phase. However, under certain conditions it can be empirically relevant. The reason is that in the absence of time delays or phase frustration, two attractively-coupled systems between which a stable phase relationship is established tend towards zero relative phase when the energy exchange rate is sufficiently large.\cite{rosenblum1996,boccaletti2002,boccaletti2018}\\
The examples which follow illustrate the potential advantages of warped phase coherence as a measure of synchronization across diverse simulated and experimental scenarios.\\
\section{R\"ossler systems}\label{rossler}
\begin{figure}
\begin{center}
\includegraphics[width=\textwidth]{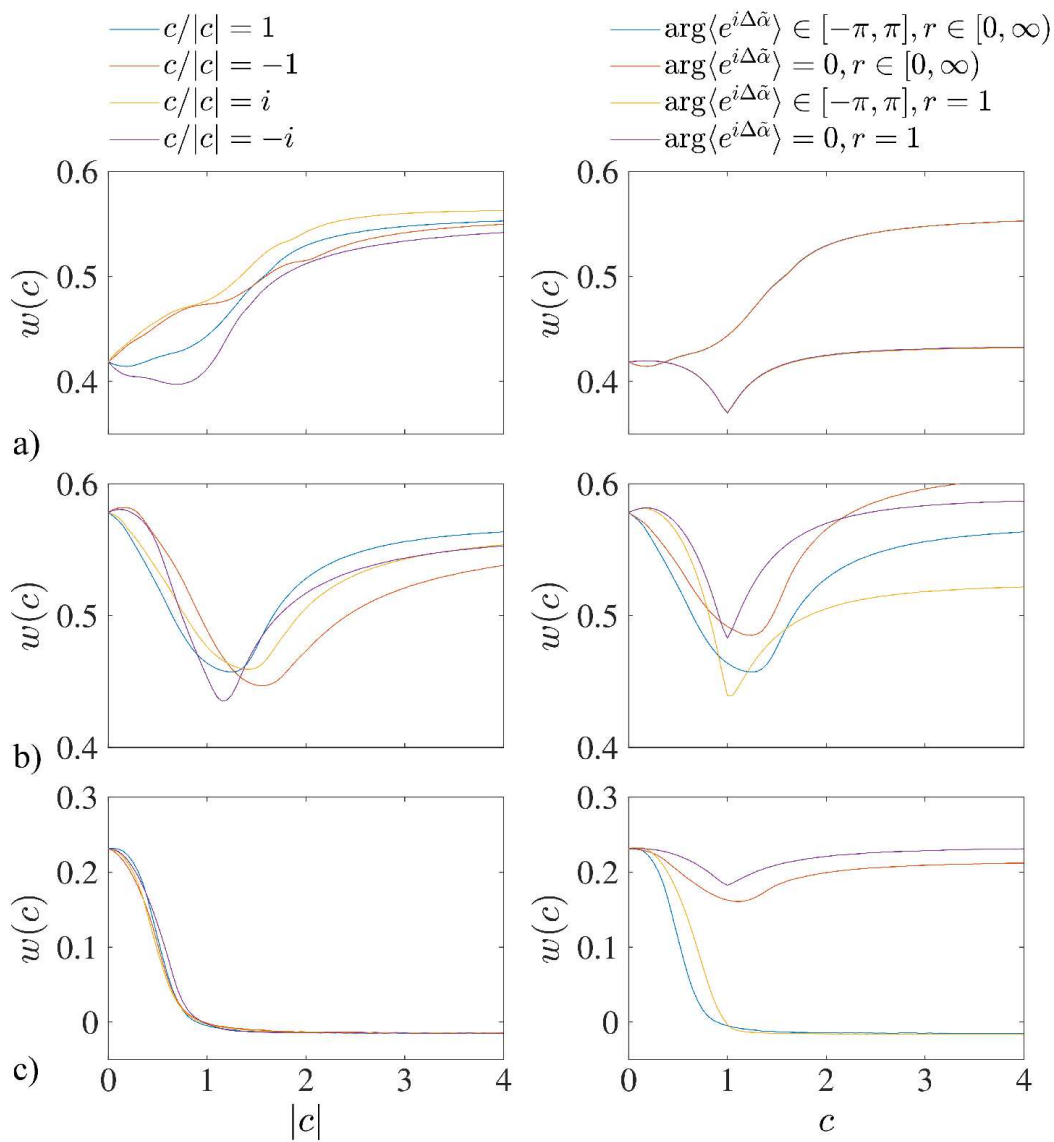}
\end{center}
\caption{Warped phase coherence $w(c)$ as a measure of synchronization between pairs of coupled R\"ossler systems, showing separately the effects of warping direction (left) and of removing amplitude and/or average relative phase (right; $r=1$, $\arg\langle e^{i\Delta\tilde{\alpha}} \rangle=0$). a), b) and c) denote different system parameter settings, see Section \ref{rossler} for description.}
\label{fig:fig6}
\end{figure}
\begin{figure}
\begin{center}
\includegraphics[width=\textwidth]{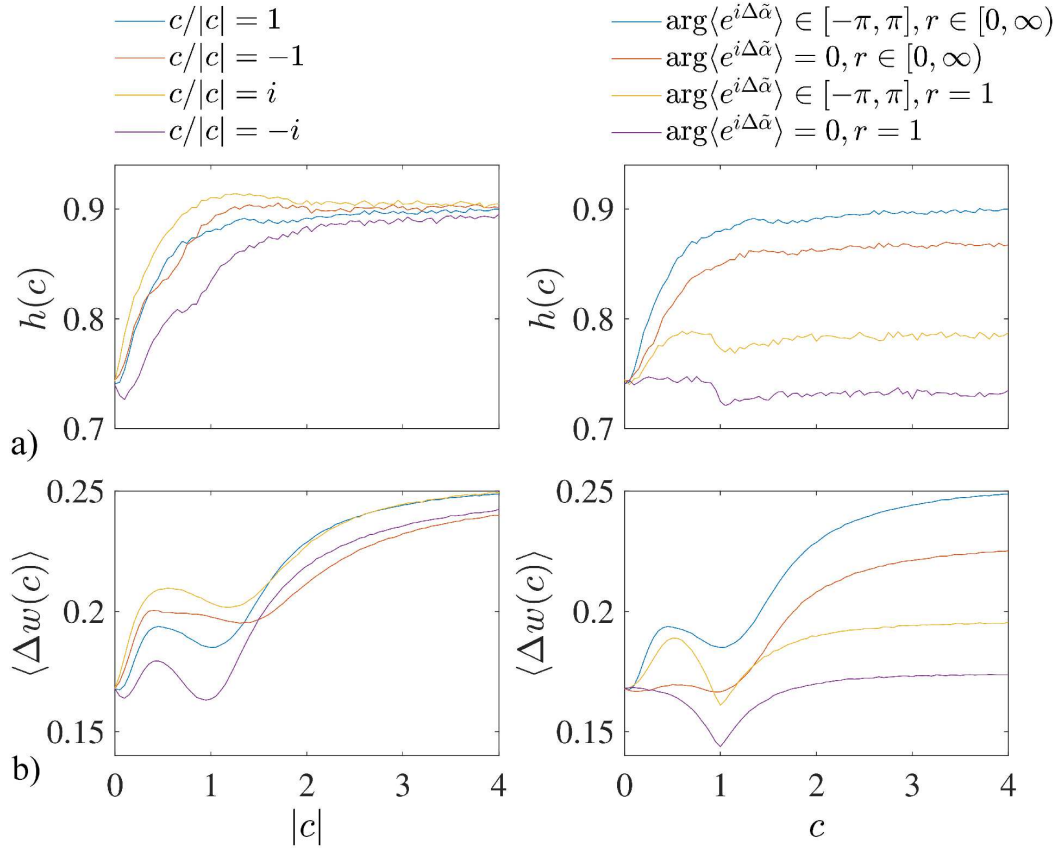}
\end{center}
\caption{Warped phase coherence as a means of inferring the structurally-coupled nodes from the generated time-series in random networks of R\"ossler systems. a) Accuracy in classifying the coupled vs. uncoupled node pairs $h(c)$, b) Corresponding average difference in coherence $\langle \Delta w(c) \rangle$.}
\label{fig:fig7}
\end{figure}
\begin{figure}
\begin{center}
\includegraphics[width=\textwidth]{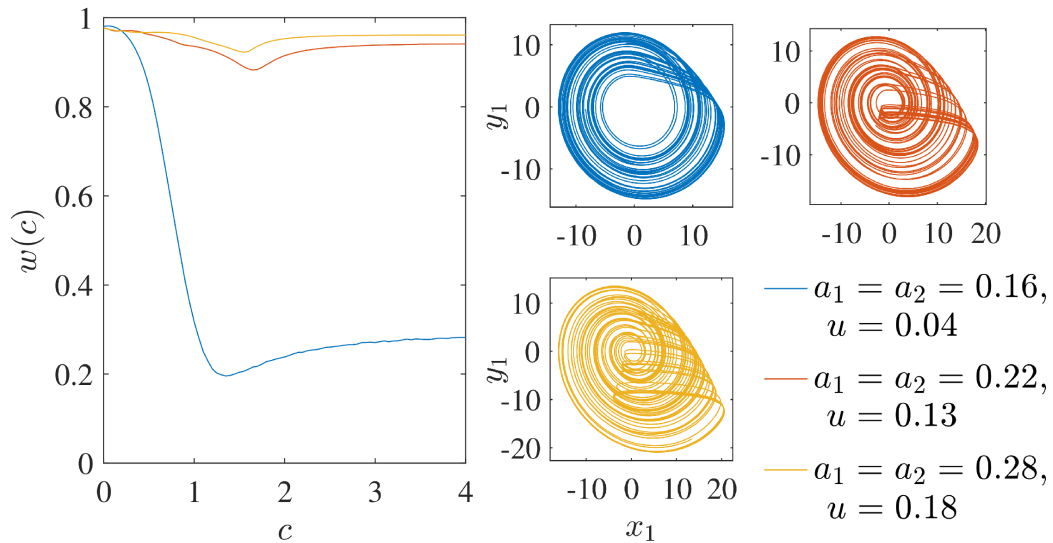}
\end{center}
\caption{Warped phase coherence $w(c)$ as a measure of synchronization (left) in an alternative set of three pairs of coupled R\"ossler systems, featuring phase-coherent and funnel attractors (right). See Section \ref{rossler} for description.}
\label{fig:fig8}
\end{figure}
To gain further insight beyond a purely stochastic situation, we next considered a representative case of synchronization of non-linear dynamics. Namely, two R\"ossler systems were diffusively coupled at the variable $x_j$, having the form
\begin{empheq}[left=\empheqlbrace]{align}
\dot{x}_{1,2}&=-y_{1,2}-z_{1,2}+u(x_{2,1}-x_{1,2})\nonumber\\
\dot{y}_{1,2}&=x_{1,2}+a_{1,2}y_{1,2}\textrm{ .}\\
\dot{z}_{1,2}&=b_{1,2}+z_{1,2}(x_{1,2}-c_{1,2})\nonumber
\end{empheq}
These were integrated by means of an explicit Runge-Kutta (4,5) formula, the Dormand-Prince pair, from $t=0$ to $t=20\times10^3$ starting from and averaging over 100 sets of identical random initial conditions with $x_1(0)=x_2(0)\in[-1,1]$, $y_1(0)=y_2(0)\in[-1,1]$ and $z_1(0)=z_2(0)\in[-1,1]$. We assessed the level of synchronization in terms of $s=x$ for three parameter settings.\\
First, we considered the case of $a_1=a_2=0.3$, $b_1=0.2$, $b_2=0.22$, $c_1=5.7$, $c_2=6.3$, $u=0.1$, which represents a moderate level of coupling and mismatching, i.e. 10\% in parameters $b$ and $c$, within a region featuring well-developed chaoticity with dense folding.\cite{rossler1976,ottbook} Despite some different fluctuations, regardless of the warping direction the coherence gradually increased from $w(0)=0.42$ to $w(4)\approx0.55$ (Fig. \ref{fig:fig6}a).\\
As shown, unlike the non-warped value $w(0)$, for $|c|>0$ $w(c)$ depends on both the amplitude fluctuations and the average relative phase between the initial signals. To better understand the consequences of warping, it is useful to separate these effects by removing either or both elements from the initial signals. That is, on the one hand setting $r=1$, i.e. considering $\psi_j(t)=e^{i\alpha_j(t)}$, and on the other hand setting $\arg\langle e^{i\Delta\tilde{\alpha}_{12}} \rangle=0$, that is, considering $\psi_j(t)=r_j(t)e^{i\tilde{\alpha}_j(t)}$, where $\tilde{\alpha}_1=\alpha_1-\arg\langle e^{i\Delta\alpha_{12}} \rangle/2$ and $\tilde{\alpha}_2=\alpha_2-\arg\langle e^{i\Delta\alpha_{21}} \rangle/2$. Doing so revealed that, in this case, the increase in $w(c)$ reflects amplitude synchronization: it is insensitive to the average relative phase but abolished by removing the amplitude component, leaving only a dip around $|c|=1$, possibly related to the ``plucking'' effect described previously (Fig. \ref{fig:fig6}a).\\
Second, we changed  $a_1=a_2=0.25$ and reduced $u=0.045$, yielding an overall comparable level of entrainment as the previous example but in a region with sparser folding.\cite{rossler1976,ottbook} In this case, the trend of $w(c)$ was considerably different, with $w(0)\approx w(4)$ and, in lieu of the increase, a dip around $c\approx1$ was observed for all settings. This result serves to illustrate that this measure is sensitive to the system dynamics, responding to the warping in a different manner depending on the level of folding and the envelope features (Fig. \ref{fig:fig6}b).\\
Third, we considered an uncoupled case with weak mismatching, setting $a_1=a_2=0.2$, $b_1=0.2$, $b_2=0.205$, $c_1=c_2=5.7$, $u=0$. Clearly, in this case $\mathbb{E}[w(c)]=0$, however one may observe $w(c)>0$ as a consequence of the finite simulation time. Whereas $w(0)=0.23$, for $|c|\ge1$ we observed $w(c)\approx0$, hallmarking an improved representation of the actual lack of entrainment in this configuration. In contrast with the first case, here the effect reflected an increased sensitivity to the non-null average relative phase, considering which enhanced detection of the actual independence between the two systems (Fig. \ref{fig:fig6}c).\\
We subsequently considered 300 sparse random networks of $n=30$ R\"ossler systems diffusively coupled at variable $x_j$ according to random binary undirected graphs $G$ obtained from an Erd\"os-R\'enyi $G(n,p)$ model having $p=0.05$. Denoting with $g_{jk}$ the entries of the corresponding adjacency matrix, for each system $j=1\ldots n$ one has
\begin{empheq}[left=\empheqlbrace]{align}
\dot{x}_j&=-y_j-z_j+u\sum_{k=1}^ng_{jk}(x_k-x_j)\nonumber\\
\dot{y}_j&=x_j+ay_j\textrm{ .}\\
\dot{z}_j&=b_j+z_j(x_j-c_j)\nonumber
\end{empheq}
Here, we set $a=0.3$ and $u=0.05$, and the other parameters were drawn randomly from $b_j\in[0.19,0.21]$ and $c_j\in[5.4,6.0]$. Integration was performed from $t=0$ to $t=6\times10^3$ ($\approx\ 850$ cycles), in this case starting from non-identical random initial conditions $x_j(0)\in[-1,1]$, $y_j(0)\in[-1,1]$ and $z_j(0)\in[-1,1]$. These simulations addressed the identification of asynchrony vs. weak synchronization between the coupled node pairs, among which $\langle w(0)\rangle=0.2$: this represents one of the scenarios previously evaluated while inferring connectivity from time-series in Kuramoto and R\"ossler networks\cite{tirabassi2015}. Denser networks with stronger couplings, wherein physical links may need to be distinguished from emergent forms of preferential entrainment, were not considered here.\cite{timme2014,rubido2015}\\
For each network, given $s_j=x_j$ the synchronization matrices $w_{jk}(c)$ were calculated with Eq. (9) for each setting of $c$ (here, the non-normalized values $\hat{w}_{jk}(c)$ were not considered). The corresponding values were then sorted and the indices $(j^{(c)}_q,k^{(c)}_{q})$ denoting the $q=1\ldots m$ most intensely synchronized node pairs were recorded, where $m$ denotes the number of edges in the graph $G$. The accuracy $h(c)\in[0,1]$ in inferring a predetermined number of structurally-coupled node pairs from synchronization of the generated time-series was thereafter determined as
\begin{equation}
h(c)=\frac{1}{m}\sum_{q=1}^mg_{j^{(c)}_qk^{(c)}_{q}}\textrm{ ,}
\end{equation}
and averaged over all the networks. The accuracy steadily increased with $|c|$ from the level of observed for the non-warped value, $h(0)=0.74$, towards $h(4)\approx0.90$. Notably, in this case, both amplitude and average relative phase information were relevant, and removing either reduced the accuracy, respectively down to $h(4)=0.79$ and $h(4)=0.87$. Since the amplitude fluctuations were slow compared to the time-scale of oscillation cycles, the warping direction had a limited effect (Fig. \ref{fig:fig7}a). For $|c|>,1$, $h(c)$ settled on a plateau close to the accuracy level observed for the correlation coefficient computed on the raw signals $s_j$, namely $h=0.92$. By comparison, the discriminatory ability of synchronization of the amplitude envelopes $r_j$ devoid of phase information was considerably lower, namely $h=0.70$ (data not shown). However, even within the plateau region, the average difference in synchronization between the structurally-coupled and the uncoupled nodes $\langle \Delta w(c) \rangle$ continued increasing (Fig. \ref{fig:fig7}b).\\
The sensitivity of warped phase coherence to the system dynamics can be further illustrated in a slightly different scenario, derived from a study wherein three qualitatively-different types of transitions to phase synchronization were observed.\cite{osipov2003} Namely, let us consider two coupled R\"ossler systems
\begin{empheq}[left=\empheqlbrace]{align}
\dot{x}_{1,2}&=-\omega_{1,2}y_{1,2}-z_{1,2}\nonumber\\
\dot{y}_{1,2}&=\omega_{1,2}x_{1,2}+a_{1,2}y_{1,2}+u(y_{2,1}-y_{1,2})\textrm{ ,}\\
\dot{z}_{1,2}&=b_{1,2}+z_{1,2}(x_{1,2}-c_{1,2})\nonumber
\end{empheq}
where $\omega_1=0.98$, $\omega_2=1.02$, $b_1=b_2=0.1$, $c_1=c_2=8.5$ and where parameter $a$ controls the attractor topology. For relatively low values, e.g. $a_1=a_2=0.16$, the attractors are phase-coherent and phase synchronization ensues already for coupling strengths at which the amplitudes remain only weakly correlated. For intermediate and high values, e.g. $a_1=a_2=0.22\textrm{ and }0.28$, the attractors become funnel: in the former case, phase synchronization sets in via an interior crisis, whereas in the latter case, it appears as a manifestation of a generalized relationship. Setting the coupling strength to values suitable for observing a similar level of phase locking $w(0)$, respectively $u=0.04\textrm{, }0.13\textrm{ and }0.18$, revealed a marked difference in the effect of increasing the warping level (Fig. \ref{fig:fig8}). For the phase-coherent attractors, a sharp drop was observed from $w(0)=0.98$ to $w(4)\approx0.26$, whereas for the two settings yielding funnel attractors, one has $w(0)\approx w(4)\approx0.95$. While a detailed analysis is beyond the scope of this work, this is in line with the previously-established facts. In the former case, phase diffusion was weak, so phase synchronization ensued quickly but in the absence of coherent amplitude fluctuations; by contrast, in the latter two cases, it was the manifestation of a deeper level of generalized entrainment.\cite{osipov2003} This was confirmed by considering different coupling strengths, at which the level of entrainment accordingly changed but the qualitative difference in the effect of the warping parameter remained: for the phase-coherent attractors, a drop is always observed, until the coupling is strong enough to entrain the amplitudes, whereas for the funnel attractors, the effect of warping remains always more constrained.
\section{Single-transistor oscillators}\label{transistor}
\begin{figure}
\begin{center}
\includegraphics[width=\textwidth]{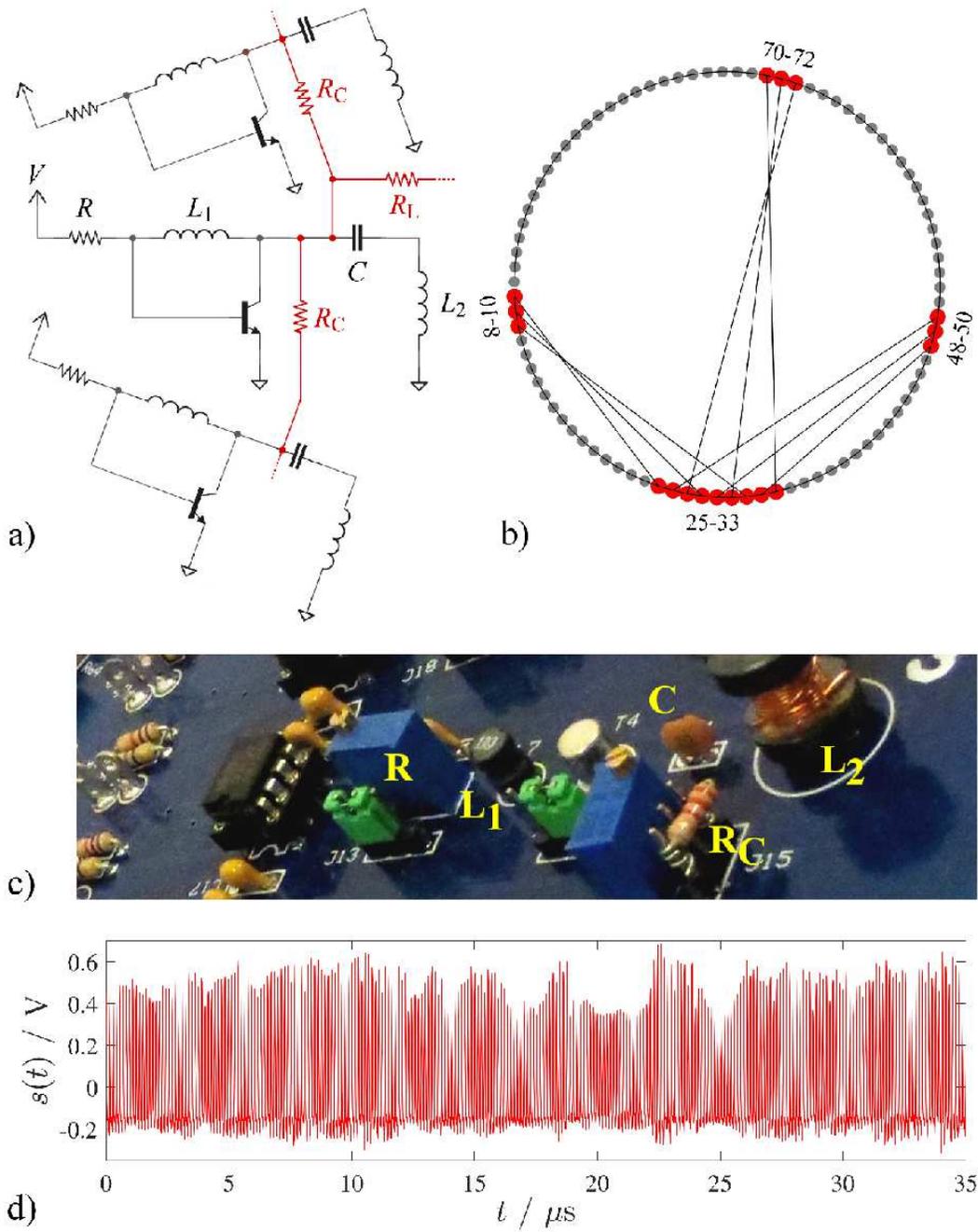}
\end{center}
\caption{Single-transistor oscillator network. a) Ring fragment showing three adjacent oscillators, wherein $R\approx1180\ \Omega$ (adjusted individually), $V=5\textrm{ V}$, $L_1=10\ \mu\textrm{H}$, $L_2=8.2\ \mu\textrm{H}$, $C=30\textrm{ pF}$, and wherein $R_\textrm{C}=750\ \Omega$ and $R_\textrm{L}=40\ \Omega$ implement, respectively, short- and long-distance links (additional 2.2 nF blocking capacitor in series with $R_\textrm{L}$ not shown). b) Network topology, with the ring segments representing ``hub regions'' in red and corresponding node numbers. c) Physical realization of an oscillator. d) Representative time-series. See Ref. (21) for a detailed description.}
\label{fig:fig9}
\end{figure}
\begin{figure*}
\begin{center}
\includegraphics[width=\textwidth]{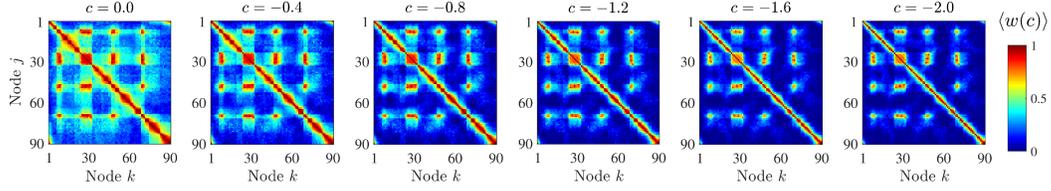}
\end{center}
\caption{Effect of increasing the warping parameter $-c$ on the spatial pattern of warped phase coherence $\langle w_{jk}(c)\rangle$ measured from the experimental network and averaged over six rotations of the link topology over the physical oscillators, as detailed in Ref. (21). Underlying adjacency matrix represented as a graph in Fig. \ref{fig:fig9}b.}
\label{fig:fig10}
\end{figure*}
\begin{figure}
\begin{center}
\includegraphics[width=\textwidth]{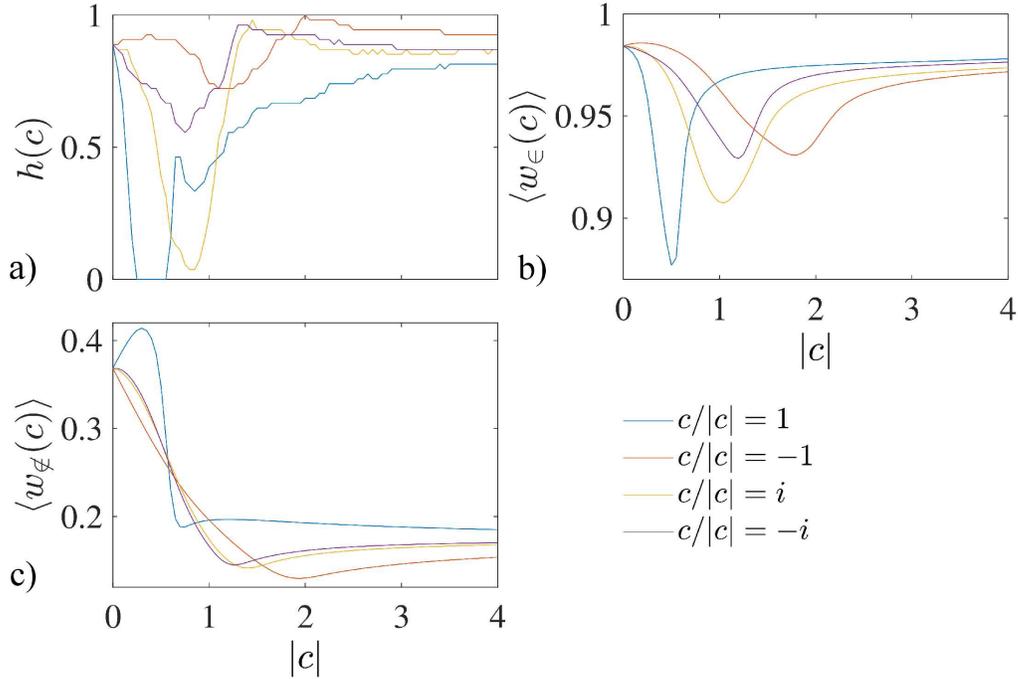}
\end{center}
\caption{Warped phase coherence as a means of inferring the stronger long-distance links from the experimental time-series. a) Accuracy in classifying the long-distance coupled vs. the other node pairs $h(c)$, b) and c) Corresponding average synchronization inside $\langle w_{\in}(c)\rangle$ and outside $\langle w_{\not\in}(c)\rangle$ the ensemble of long-distance coupled node pairs.}
\label{fig:fig11}
\end{figure}
\begin{figure}
\begin{center}
\includegraphics[width=\textwidth]{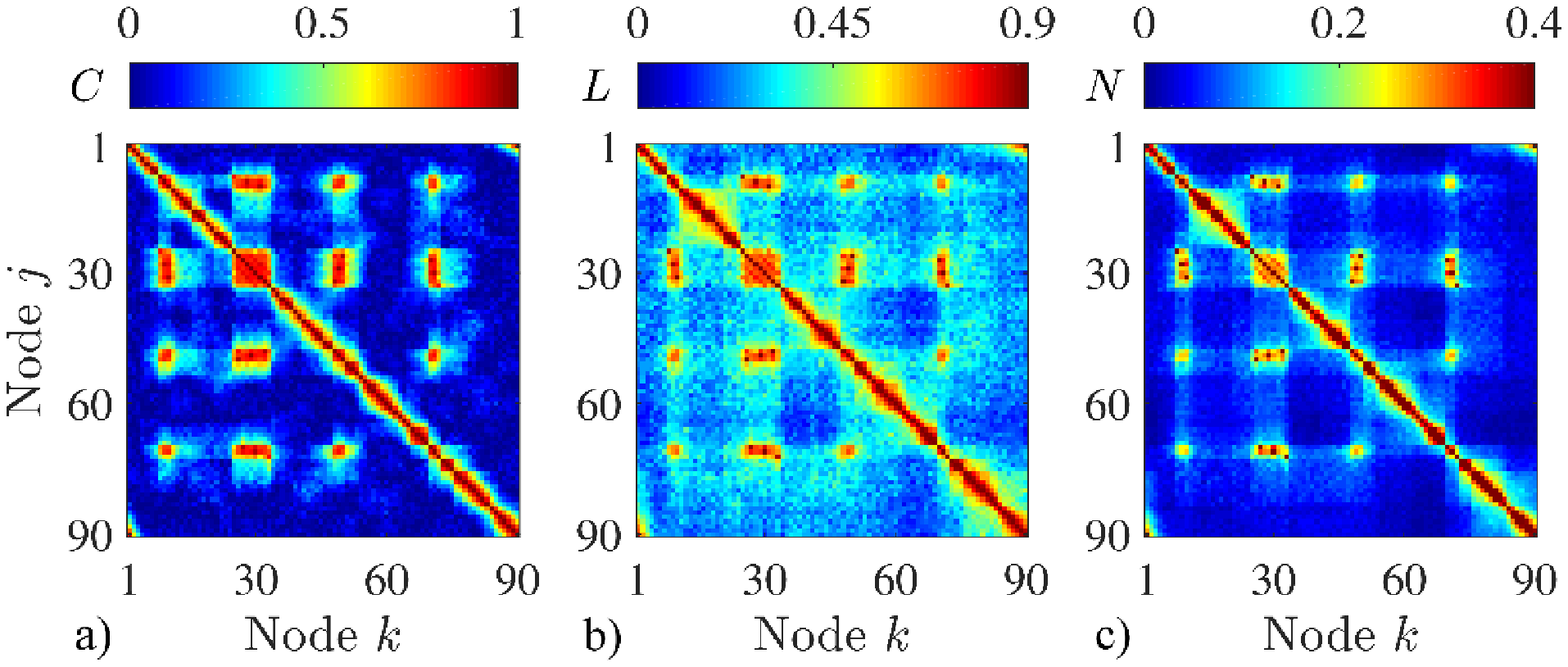}
\end{center}
\caption{Alternate synchronization measures. a) Linear correlation $C$, b) Generalized synchronization $L$ and c) Normalized mutual information $N$. See Section \ref{transistor} for description.}
\label{fig:fig12}
\end{figure}
To evaluate the proposed synchronization measure in an experimental scenario, we next considered a structured network of coupled single-transistor oscillators. Each node consists of an ``atypical'' circuit resembling a Hartley oscillator, whose autonomous dynamics are controlled via a resistor placed in series to the external DC voltage source. For certain resistance ranges, transition to chaos occurs, and manifests primarily in the form of cycle amplitude fluctuations (Fig. \ref{fig:fig9}a).\cite{minati2014} A ring comprising 90 such oscillators was studied, wherein nodes are diffusively coupled to their neighbors via resistors $R_\textrm{C}$ connected between the transistor collectors. A minority of nodes also have an additional resistor $R_\textrm{L}$ towards a distant site, delineating four segments which yield a ``toy model'' of the hub nodes in a complex network (Fig. \ref{fig:fig9}b). The system was physically realized with discrete components on a circuit board (Fig. \ref{fig:fig9}c).\cite{minati2015}\\
For the present purpose, we reanalyzed publicly available time-series recorded in a collectively-chaotic phase (50,000 points, $\approx750$ cycles; example in Fig. \ref{fig:fig9}d).\cite{repositorylink} We focused on the fact that the 9 long-distance connections were implemented through stronger couplings $R_\textrm{L}=40\ \Omega$ than the 90 short-range links $R_\textrm{C}=750\ \Omega$ realizing the ring; hence, despite them being weakened by parasitic effects, one expects stronger entrainment to be detectable.\\
Accordingly, the non-warped value $w(0)$ revealed that synchronization was elevated within the collective of interconnected nodes in the distant segments, and it noticeably diffused to the other nodes along the ring outside these ``hub regions''. Increasing the magnitude of the warping parameter, particularly along the $c/|c|=-1$ direction and for $w(c)$, had a striking ``focusing'' effect: it gradually reduced the intensity of the synchronization diffused outside these regions and attenuated the clusters spontaneously formed along the ring, delineating increasingly clearly the distantly-coupled segments (Fig. \ref{fig:fig10}).\\
Considering the same approach as in Section \ref{rossler} for identifying the nine long-distance links from the time-series, a non-monotonic effect on accuracy was noted, starting from $h(0)=0.89$, markedly decreasing and then recovering, eventually reaching $h(-2)=1$ (Fig. \ref{fig:fig11}a); an analogous difference was observed for the non-normalized values $\hat{w}(c)$ (data not shown). Underlying this observation there appeared to be two competing, qualitatively different effects recalling the simulations of R\"ossler systems in Fig. \ref{fig:fig6}b versus Fig. \ref{fig:fig6}c. 
Among the long-distance coupled nodes the situation was one of near-complete synchronization, and warping had a non-monotonic effect on the measured coherence, delineating a dip down from $\langle w_{\in}(0)\rangle=0.98$. As a consequence, for $|c|<2$ these links transiently became less clearly distinguishable from preferential entrainments along the weaker short-distance connections (Fig. \ref{fig:fig11}b). Differently from the coupled R\"ossler oscillators and the electroencephalogram, the dip location and depth markedly depended on the warping direction. The reason is that activity was spike-like, implying that the amplitudes fluctuated considerably on the temporal scale of individual cycles, hence there was considerable spectral overlap between $s_j(t)$ and $r_j(t)$. On the other hand, across the rest of the network the measured coherence steadily decreased with warping, delineating a sharp drop from $\langle w_{\not\in}(0)\rangle=0.37$ to $\langle w_{\not\in}(4)\rangle\approx0.17$ (Fig. \ref{fig:fig11}c), primarily due to increased sensitivity to the non-null average relative phase (data not shown). As a consequence, the ensemble of directly structurally-connected nodes eventually became more clearly distinguishable.\\
The relationship between phase- and amplitude-related effects can be intricate, and a proper comparison with other synchronization measures is beyond the scope of this initial study, in that it would require comprehensive consideration of diverse system dynamics, coupling strengths, time-series lengths etcetera. Nevertheless, for the only purpose of visual comparison, three other representative synchronization measures\cite{boccaletti2002} were computed for these time-series. First, linear correlation, which assumes $y(t)\propto x(t)$. Second, generalized synchronization, which assumes a possibly more complicated functional relationship between the trajectories $\mathbf{y}(t)=\mathbf{\Phi}(\mathbf{x}(t))$; this was estimated via a rank-based measure, namely the $L$-index, setting $\tau=40\textrm{ ns}$, $w=100\textrm{ ns}$, $m=5$ and $k=m+1$.\cite{chicharro2009} Third, normalized mutual information, which assumes a possibly non-linear but scalar relationship $y(t)=\Phi(x(t))$; this was estimated as $I_\textrm{XY}/\sqrt{H_\textrm{X}H_\textrm{Y}}$, where $I$ and $H$ represent, respectively, the mutual information and the Shannon entropies, calculated over $N=16$ bins over a maximum lag $d=\pm40\textrm{ ns}$.\cite{paninski2003} Taking into account the different sensitivity of these measures, also to noise, it appears evident that the resulting synchronization matrices are overall more similar to those yielded by warped phase coherence $w(c)$ with relatively large values of $c$ than to that obtained for the non-warped value $w(0)$ (Fig. \ref{fig:fig12}). This is in line with the view that warping confers sensitivity to aspects of entrainment beyond the mere phase locking, as was also indicated by the simulations of R\"ossler systems considered in Section \ref{rossler}. It should also be noted that rank-, bin- and time-lag embedding based measures are inherently poorly suited for use on short time-series epochs, such as those considered next in Section \ref{eeg} (640 vs. $>10,000$ points).\cite{chicharro2009,paninski2003,faes2008}
\section{Electroencephalogram}\label{eeg}
\begin{figure*}
\centering
\includegraphics[width=\textwidth]{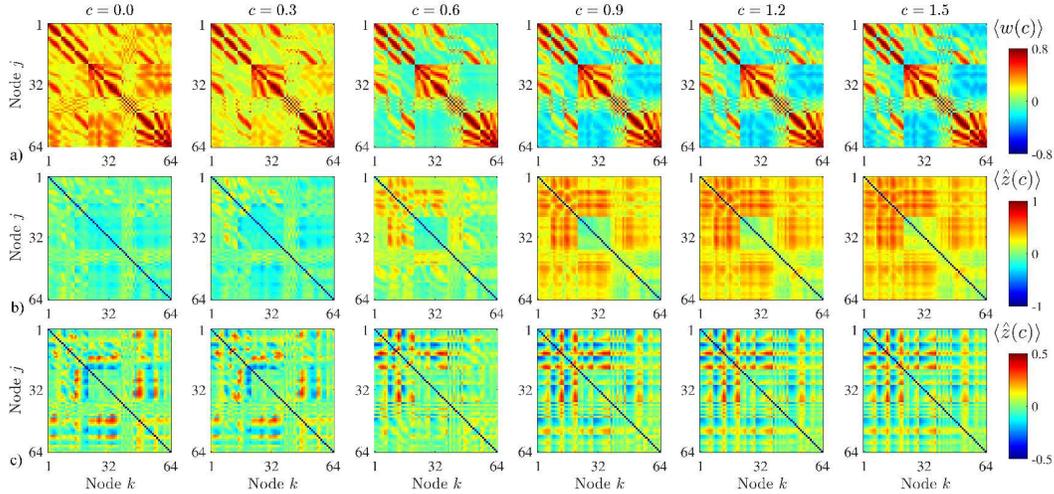}
\cprotect\caption{Effect of increasing the warping parameter $c$ on the spatial pattern of warped phase coherence across the electrodes located on different scalp regions. a) Average over all conditions and participants $\langle w_{jk}(c)\rangle$. b) Rest-activation and c) left-right imagery contrasts expressed in units of pooled standard deviation $\langle \hat{z}_{jk}(c)\rangle$. Pericentral electrodes (FCx, Cx and CPx sites): $j=1\ldots21$. Frontal electrodes (AFx, Fx and FPx sites): $j=22\ldots38$. Parietal-occipital electrodes (Px, POx and Ox sites): $j=47\ldots64$. Electrodes overlaying the sensorimotor cortex: left (C5, C3, CP5 and CP3 sites) $j=8,9,15,16$ and right (C4, C6, CP4 and CP6 sites) $j=13,14,20,21$.\label{fig:fig13}}
\end{figure*}
\begin{figure}
\centering
\includegraphics[width=0.95\textwidth]{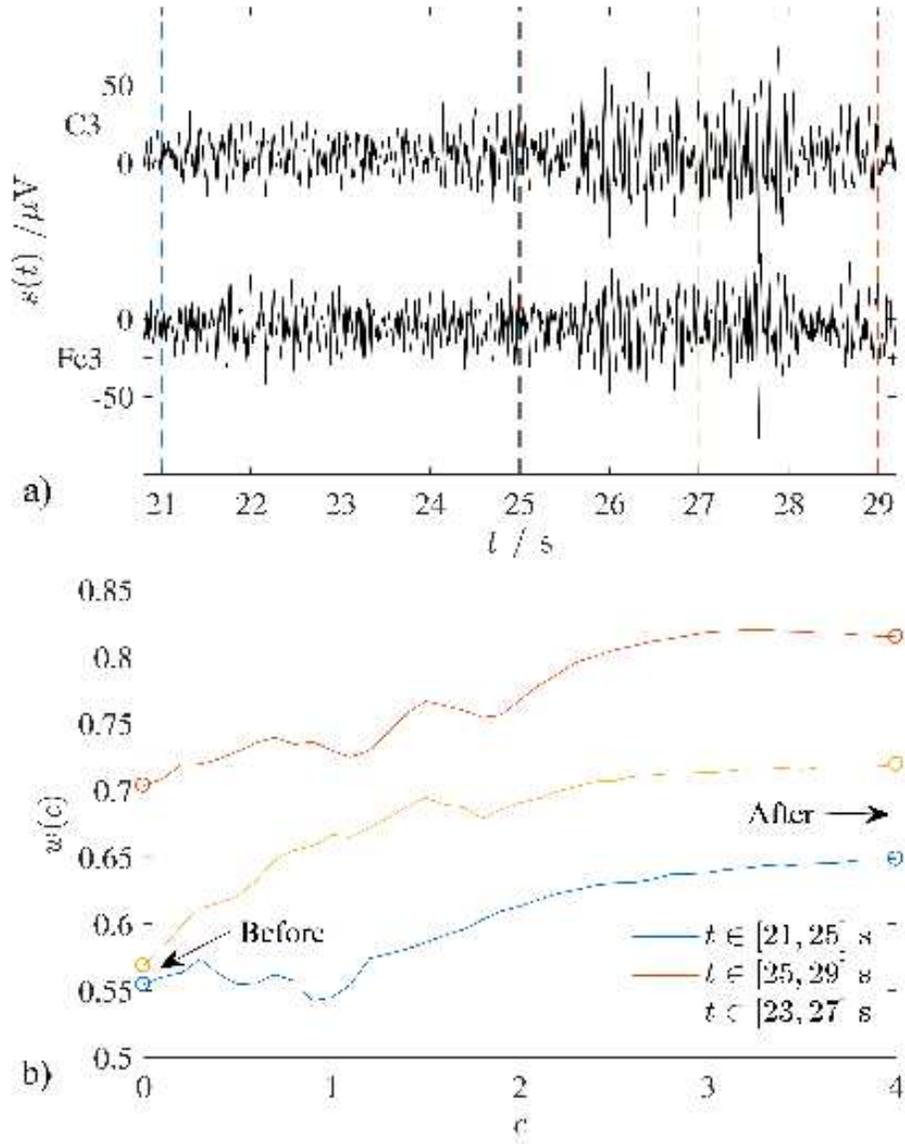}
\cprotect\caption{Representative example of the effect of warping on determining the level of synchronization between segments of two electroencephalographic signals. a) Chosen signals and corresponding time-windows for synchronization determination. b) Effect of increasing the warping parameter $c$ on warped phase coherence in the three time-windows (curves averaged over 100 evaluations of Eq. (9)). For the intermediate time-window, the measured coherence is very close to that of the left time-window before warping, but more appropriately intermediate between the left and right time-windows after warping is applied. See Section \ref{eeg} for description.\label{fig:fig14}}
\end{figure}
\begin{figure}
\centering
\includegraphics[width=\textwidth]{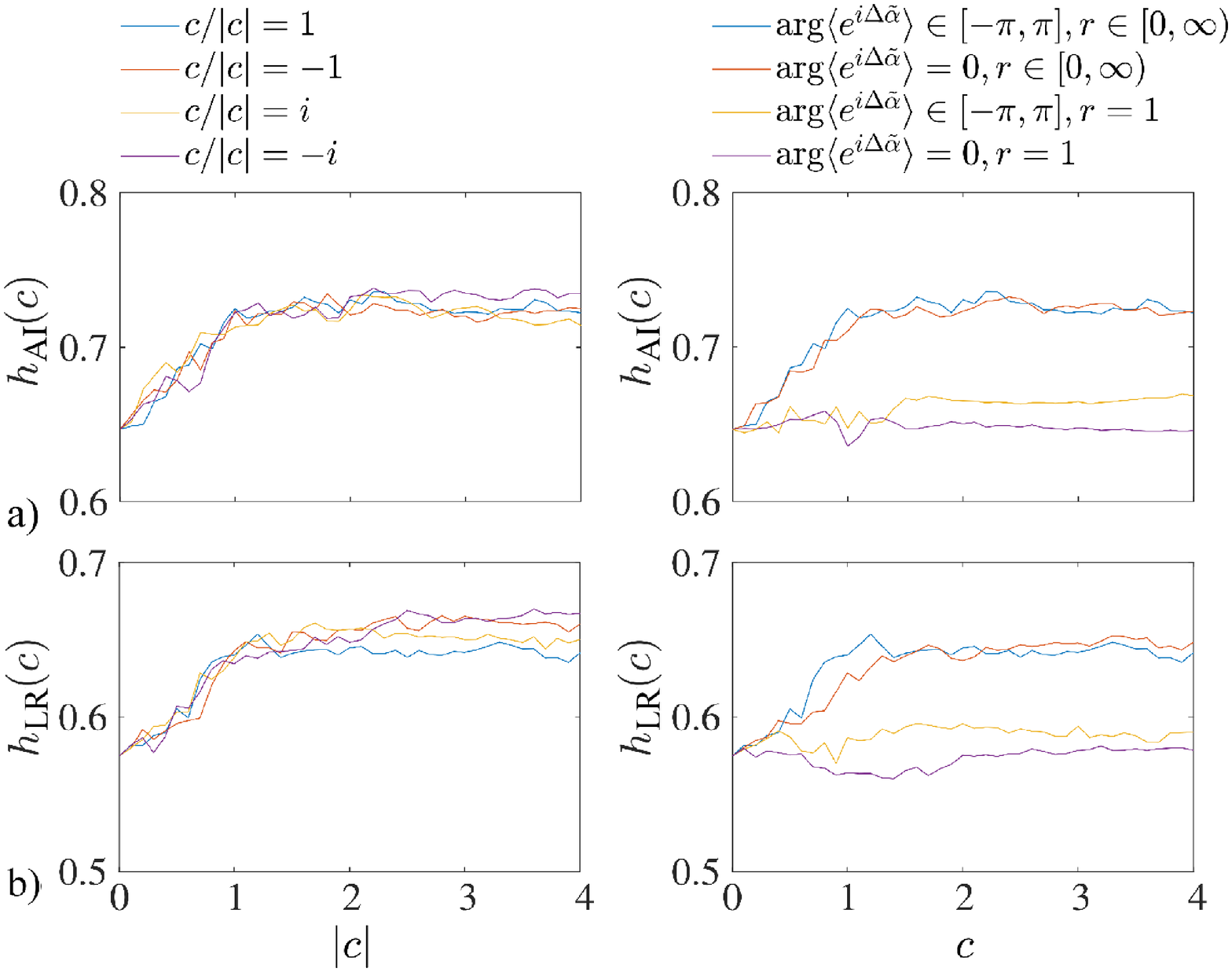}
\cprotect\caption{Warped phase coherence as a means of detecting brain states corresponding to imaginary actions. a) Accuracy in classifying epochs of activity or idleness $h_\textrm{AI}(c)$. b) Accuracy in classifying epochs of left or right hand motor imagery $h_\textrm{LR}(c)$.\label{fig:fig15}}
\end{figure}
Finally, we demonstrate the empirical value of the proposed measure for analyzing electroencephalographic signals with the purpose of decoding imaginary actions, representing a typical scenario in brain-computer interface systems.\cite{valerdi2015} This is only presented here to exemplify a hypothetical application. Explicit comparison with established approaches for constructing connectivity matrices from these signals, such as nonlinear associations\cite{chicharro2009,paninski2003}, Granger causality\cite{faes2008} and wavelet bicoherence\cite{li2009}, is mandatory but left for future work.\\
In electroencephalographic signals, determinism is considerably weaker compared to the cases considered thus far, and there may be contamination by non-neural noise sources. We re-analyzed a publicly-available dataset of 106 recordings from healthy volunteers, and we focused on a task alternating idle wakefulness with actively imagining to repeatedly open and close the right or left fist (respectively 45, 22 and 23 trials). This dataset yielded 4 s-long epochs sampled at 160 Hz (640 points) over $n=64$ channels, from electrodes laid out onto the scalp according to the international 10-10 system.\cite{goldberger2000,schalk2004,repositorylink2}\\
Given the dense electrode montage, spatial filtering was firstly applied by subtracting to each signal $s_j(t)$ with $j=1\ldots n$ the average reference channel $\langle s(t) \rangle=\frac{1}{n}\sum_{j=1}^ns_j(t)$. Then, temporal filtering was performed over the frequency range from 7 to 30 Hz using a second-order band-pass Butterworth filter, which preserved activity in the $\alpha,\mu$- and $\beta$-bands. These bands index different aspects of movement planning and execution, and allow attenuating any slow drifts and electromyographic contamination.\cite{neuper2001,aghaei2016,chella2016} To prevent selection bias, we refrained from manually inspecting and rejecting any recording. Following calculation of the analytic signal with Eq. (1), epoching was performed. The warped phase coherence matrix was thereafter obtained with Eq. (8), separately for each epoch. Brain states were binarily classified at individual-participant level, separately for each setting of the warping parameter $c$. Sparse logistic regression with variational approximation was performed over all $(n^2-n)/2$ synchronization values, and the binary contrasts active-idle and left-right hand were considered. The corresponding decoding accuracy $h(c)$ was quantified via the pair-wise leave-one-out approach over 20 random combinations.\cite{yamashita2008}\\
Consideration of the non-warped value $\langle w(0)\rangle$, averaged over all epochs and participants, revealed strong synchronization within the pericentral electrodes, especially between adjacent, ipsilateral subsets (diagonal lines), as well as within and between the parietal-occipital and the frontal electrodes. Particularly over the range $c\in[0,1.5]$, increasing the warping parameter $c$ had a prominent effect, recalling the observations on the electronic oscillators: spatial structure became considerably more evident and the value distribution became wider, expanding from $\langle w(0)\rangle\in[0.1,0.8]$ to $\langle w(1.5)\rangle\in[-0.4,0.8]$. The average coherence decreased and preferential entrainment between ipsilaterial electrodes over the central-parietal region consequently became emphasized (Fig. \ref{fig:fig13}a). This trend, which was similar for the non-normalized values $\langle \hat{w}(c)\rangle$, reflected amplitude-related effects as well as average relative phase. It should be noted that, even though in principle volume conduction could pose a problem since warping leads to higher coherence values for small phase lags, it remains unsettled whether zero-lag connectivity contains useful information or not.\cite{jian2017,bruna2018} On another note, unlike for the electronic oscillators, here the spectra of $s_j(t)$ and $r_j(t)$ had limited overlap as only the latter were $1/f$-like, hence the warping direction had limited influence (data not shown).\\
To aid understanding one of the possible effects of warping on determining the level of synchronization between electroencephalographic signals, it is helpful to consider a representative example, consisting of signals acquired over two pericentral sites. In particular, let us consider a segment: towards the beginning the temporal activity appears relatively disordered, low-amplitude and unsynchronized, and towards the end a volley (or ``spindle'') of large-amplitude, more coherent $\mu$-band oscillations occurs (Fig. \ref{fig:fig14}a). It is helpful to consider warped phase coherence within three overlapping 4 s-long time-windows: two covering only the baseline activity and only the large-amplitude oscillations (respectively, until and after $t=25\text{ s}$), and one covering an intermediate interval wherein both types of activity are found. For this latter time-window, in the absence of warping the measured level of synchronization was very close to that of the baseline activity: it failed to represent the fact that its second half contains the initial part of the large-amplitude volley (i.e., $w(0)\approx0.56$). As the warping parameter was elevated, the measured coherence increased up to a level approximately half-way between the two non-overlapping time-windows (e.g. $w(2)=0.69\textrm{ vs. }0.61, 0.77$): this more appropriately represented the presence of a combination of both types of activity within this intermediate window (Fig. \ref{fig:fig14}b). Volleys of coherent $\mu$-band oscillations are knowingly generated in relation to motor tasks by event-related synchronization phenomena, and, as introduced in Section \ref{formulation}, their detection can be enhanced by warping primarily as a reflection of their larger amplitude.\cite{mcfarland2000,neuper2001}\\
The task-related contrasts were extracted from the non-normalized values $\langle \hat{w}(c)\rangle$, else classification accuracy would be reduced. Notably, task-related differences were also attenuated when removing the amplitude differences between channels through individually setting $\langle |\psi_j| \rangle=1$ (data not shown). The active-idle difference normalized in units of pooled standard deviation $\hat{z}_{ij}(c)$ became visibly stronger with increasing $c$: during task performance, coherence markedly emerged within the pericentral electrodes, and between them and the rest of the network (Fig. \ref{fig:fig13}b). As expected, the left-right imagery separation was more subtle, but there was a clearly-identifiable pattern wherein warping attenuated differences elsewhere on the scalp. Simultaneously, it enhanced a modulation of the level of synchronization between all pericentral electrodes and specifically those directly overlying the sensorimotor cortex. This yielded a lateralized change in sign, which plausibly reflected event-related synchronization/desynchronization ipsilaterally vs. contralaterally to the imaginary movement (Fig. \ref{fig:fig13}c).\cite{mcfarland2000,neuper2001}\\
As the warping parameter was elevated, the recognition accuracy steadily improved, reaching a plateau for $|c|\ge2$. For the active-idle contrast, the accuracy increased from $h_\textrm{AI}(0)=0.65\pm0.12$ to $h_\textrm{AI}(2)=0.73\pm0.12$ ($\mu\pm\sigma$; Fig. \ref{fig:fig15}a). For the left-right hand contrast, it increased from $h_\textrm{LR}(0)=0.57\pm0.13$ to $h_\textrm{LR}(2)=0.64\pm0.15$ (Fig. \ref{fig:fig15}b). For both contrasts, the effect of warping appeared more closely related to the amplitude fluctuations than to the average phases between electrodes. In line with other works, considerable inter-individual variability was present, plausibly reflecting issues related to task performance which, being the task covert, could not be ascertained behaviorally.\cite{ahn2015} At group level, both comparisons were strongly statistically significant (paired t-test $p<0.001$), and there was no effect of the warping direction ($p>0.1$). The number of features selected by the sparse regression decreased for the first contrast ($52\pm18$ vs. $48\pm22$, $p=0.03$) but not for the second one ($51\pm13$).\\
Confirming the validity of the results, no differences were found comparing the odd with the even epochs of idleness, and all effects were obliterated by phase randomization preserving both auto- and cross-correlations (multivariate surrogates, $p>0.8$).\cite{schreiber1996} Discarding the amplitude information by assuming $r=1$, the accuracy after warping was substantially reduced ($p<0.001$). Nevertheless, a marginal improvement with respect to the non-warped value remained visible for both contrasts ($p=0.05$), reflecting sensitivity to the average relative phases. Accordingly, the accuracy obtained with warped phase coherence significantly exceeded that yielded by cross-correlation on $s_j(t)$ ($p=0.03$, data not shown). Repeating the analyses limited to the $\alpha,\mu$-band activity (7 to 13 Hz), the accuracy of active-idle classification dropped significantly ($p<0.001$), but there was no effect on left-right hand classification performance ($p=0.3$). Repeating them again for the $\beta$-band activity (13 to 30 Hz), the accuracy of warped phase coherence in both contrasts was unaffected ($p>0.4$).\cite{mcfarland2000}\\
Lastly, we considered two other behavioral tasks from the same participants and experimental sessions, with an identical blocked design. Firstly, the same motor task but involving actual hand movement. Also in this case, for the active-idle contrast the accuracy increased from $h(0)=0.75\pm0.13$ to $h(2)=0.82\pm0.10$ and for the left-right hand contrast it increased from $h(0)=0.60\pm0.12$ to $h(2)=0.68\pm0.13$ ($p<0.001$). Secondly, a different task comparing simultaneous imaginary movement of both fists or feet. Again, for the active-idle contrast the accuracy increased from $h(0)=0.65\pm0.11$ to $h(2)=0.71\pm0.11$ and for the fists-feet contrast it increased from $h(0)=0.62\pm0.16$ to $h(2)=0.67\pm0.15$ ($p<0.001$).
\section{Conclusions}\label{concleeg}
Warping yields sensitivity to amplitude fluctuations and differences in the average phase. On the one hand, this complicates the physical interpretation of phase coherence, since one loses its intuitive meaning associated with Kuramoto-like networks. As warping is increased, the measure becomes more and more distant from representing phase locking, even reminiscent of linear correlation. On the other hand, warping seems to have some empirical value for studying collective dynamics under partial synchronization across diverse systems having free amplitude. This stems from the fact that, in non-linear systems, a stable phase relationship is only one of the hallmarks of entrainment. Knowingly, as the coupling between two non-identical oscillators is strengthened, also amplitude fluctuations become increasingly coherent and, in the absence of frustration or other obstacles, lags are eventually overcome.\cite{rosenblum1996,boccaletti2002,boccaletti2018}\\
Given a certain level of phase entrainment, $w(c)$ may increase, decrease or fluctuate with the level of warping $|c|$; as illustrated by the given examples, these denote qualitatively different situations. As illustrated by the cases of coupled R\"ossler systems, due to dependence on dynamics it is not straightforward to compare warped phase coherence measurements between different systems. However, it was shown that, within a given network, this measure can aid the identification of which node pairs are most strongly entrained, surpassing both the phase locking value and cross-correlation. In particular, the consistent advantage observed when setting $c\ge2$ (assuming $\langle r \rangle$=1), plausibly related to the ``twisting'' effect of warping and consequentially increased amplitude sensitivity, appears noteworthy.\\
Even though in this exploratory study it is impossible to be prescriptive, a warping with $c\approx2$ would appear to be consistently appropriate, even across the rather diverse systems that were considered. For smaller values, a larger dependence on the level of warping may remain, whereas for larger values, $\hat{w}_{jk}(c)$ in Eq. (8) and $\hat{w}^\prime_{jk}(c)$ in Eq. (10) may become undesirably close to unity, in turn affecting the accuracy of calculating $w_{jk}(c)$ in Eq. (9).\\
More generally, the impact of the phase and amplitude distributions and the assumptions underlying and the implications of the normalization step require further elucidation (i.e., effects visible on $w_{jk}(c)$ vs. $\hat{w}_{jk}(c)$). Furthermore, systematic comparison with different synchronization measures, in particular other variants of the phase locking value, is necessary.\cite{kovach2017,lepage2017,bruna2018} With reference to aiding the inference of structural couplings, confirmation over diverse coupling strengths, more complex and denser topologies is required.\cite{tirabassi2015,timme2014,rubido2015} As regards the electroencephalogram, the influences of volume conduction and reference choice require additional analysis. In particular, one needs to consider the sensitivity to average phase differences and its impact on the determination of synchronization between scalp sites at which a given activity component is recorded with opposite polarity.\cite{chella2016,jian2017,bruna2018}\\
In any case, the observed improvement in the classification accuracy for imaginary actions, on average $\approx8\%$ and considerably higher for certain participants, appears potentially clinically-relevant. It is in line with that yielded by other algorithmic enhancements vastly more demanding to deploy, such as transforming the signals from sensor to source space via inverse modeling.\cite{ahn2015,yoshimura2016} Furthermore, the accuracy was maximized when including the entire $\alpha,\mu$- and $\beta$-bands, thereby alleviating the need for individually-adjusted frequency filter settings.\cite{neuper2001,aghaei2016} It remains to be seen whether a similar improvement could be replicated in other datasets, obtained when applying this measure in source space, and generalized in the context of different sensorimotor and cognitive tasks. As indicated above, it also remains to be determined how the present measure compares to more consolidated alternatives, particularly in the context of determining connectivity over short signal segments.\cite{chicharro2009,paninski2003,faes2008,li2009}\\
Nevertheless, consistent improvements were seen across three different motor tasks. Hence, we posit that inclusion of warping in brain-computer and brain-machine interface systems based on the phase locking value and similar indices could be an immediate means of enhancing their performance.
\section*{ACKNOWLEDGMENTS}
This work was supported in part by JSPS under Grant KAKENHI 26112004 and Grant KAKENHI 17H05903. In addition to employment by the Polish Academy of Sciences, Krak\'ow, Poland, L. Minati gratefully acknowledges funding by the World Research Hub Initiative, Institute of Innovative Research, Tokyo Institute of Technology, Tokyo, Japan. The authors are grateful to G. Schalk, W.A. Sarnacki, A. Joshi, D.J. McFarland and J.R. Wolpaw for collecting motor imagery EEG data and making them publicly available on PhysioNet.org. Their work was supported by grants from the NIH/NIBIB (EB006356 GS, EB00856 JRW and GS) and PhysioNet is supported by the NIH/NIGMS/NIBIB (2R01GM104987-09).\\

\end{document}